\documentstyle[aps,prl,amssymb,epsf,multicol]{revtex}
\renewcommand{\narrowtext}{\begin{multicols}{2} \global\columnwidth20.5pc}
\renewcommand{\widetext}{\end{multicols} \global\columnwidth42.5pc}

\begin{document}

\newcommand{\newc}{\newcommand}

\newc{\be}{\begin{equation}}
\newc{\ee}{\end{equation}}
\newc{\ba}{\begin{eqnarray}}
\newc{\ea}{\end{eqnarray}}
\newc{\bea}{\begin{eqnarray*}}
\newc{\eea}{\end{eqnarray*}}
\newc{\D}{\partial}
\newc{\ie}{{\it i.e.} }
\newc{\eg}{{\it e.g.} }
\newc{\etc}{{\it etc.} }
\newc{\etal}{{\it et al.}}
\newcommand{\nn}{\nonumber}

\newc{\ra}{\rightarrow}
\newc{\lra}{\leftrightarrow}
\newc{\no}{Nielsen-Olesen }
\newc{\lsim}{\buildrel{<}\over{\sim}}
\newc{\gsim}{\buildrel{>}\over{\sim}}

\title{Semitopological Q-Rings\footnote{Contributed to the 4th Workshop "What Comes Beyond The Standard Model", Bled,
Slovenia, July 17-27 2001 in honor of Holger Beck Nielsen's 60th 
Birthday.} }
\author{Minos Axenides$^a$}
\address{$^a$ Institute of Nuclear Physics, N.C.S.R. Demokritos,  153
10, Athens, Greece\\ e-mail: axenides@mail.demokritos.gr } 

\date{\today}
\maketitle

\begin{abstract}
Semitopological Vortices (Q-Rings) are identified to be classical 
soliton configurations whose stability is attributed to both 
topological and nontopological charges. We discuss some recent 
work on the simplest possible realization of such a configuration 
in a scalar field theory with an unbroken $U(1)$ global symmetry. 
We show that Q-Rings correspond to local minima of the energy, 
exhibit numerical solutions of their field configurations and 
derive virial theorems demonstrating their stability.  

\end{abstract}

\narrowtext

As we celebrate the 60th birthday of Holger Bech Nielsen we can 
without doubt assess his contributions to the development of the 
theory of strings and vortices to  bear the strongest possible 
impact. Indeed  his early work on the development of multiparticle 
dual models\cite{KN1,KN2,KN3} was soon after followed by the 
introduction of the string picture in the study of strong 
interaction physics \cite{HBN}. At the time it improved 
tremedously our physical understanding of dual models\cite{PF}. 
The string concept, of course, was bound to become much more 
useful in the unification of particle interactions with gravity. 
Aside from Holger's contribution to the development of the "string 
idea" he much later provided the first covariant formulation of a 
vortex in a theory with spontaneously broken abelian gauge 
symmetry \cite{NO}. The stability of such a gauged vortex is due 
the presence of a topological charge. The cosmic role of such 
topological defects in the phase transitions of the early universe 
has been important. It is in the spirit of this line work of 
Holger's that we will present a novel class of vortex like 
configurations that share some of the properties of topological 
solitons as well as those that are non-topological in 
character.Hence their identification as semitopological. The work 
was done in collaboration with E.G.Floratos,S.Komineas and 
L.Perivolaropoulos\cite{AFKP}  

 Non-topological solitons (Q balls)  are localized time dependent field 
configurations with a rotating internal phase and their stability 
is due to the conservation of a Noether charge $Q$\cite{c85}. They 
have been studied extensively in the literature in one, two and 
three dimensions\cite{lp92}. In three dimensions, the only 
localized, stable configurations of this type have been assumed to 
be of spherical symmetry hence the name Q balls. The 
generalization of two dimensional (planar) Q balls to three 
dimensional Q strings leads to loops which are unstable due to 
tension. Closed strings of this type are naturally produced during 
the collisions of spherical Q balls and have been seen to be 
unstable towards collapse due to their tension\cite{bs00,lpwww}. 

There is a simple mechanism however that can stabilize these 
closed loops. It is based on the introduction of an additional 
phase on the scalar field that twists by $2\pi N$ as the length of 
the loop is scanned. This phase introduces additional pressure 
terms in the energy that can balance the tension and lead to a 
stabilized configuration, the {\it Q ring}. This type of pressure 
is analogous to the pressure of the superconducting string 
loops\cite{Witten:1985eb} (also called `springs'\cite{hht88}). In 
fact it will be shown that Q rings carry both Noether charge and 
Noether current and in that sense they are also superconducting. 
However they also differ in many ways from superconducting 
strings. Q rings do not carry two topological invariants like 
superconducting strings but only one: the winding $N$ of the phase 
along the Q ring. Their metastability is due not only to the 
topological twist conservation but also  due to the conservation 
of the Noether charge as in the case of ordinary Q balls. Due to 
this combination of topological with non-topological invariants Q 
rings may be viewed as semitopological defects. In what follows we 
demonstrate the existence and metastability of Q rings in the 
context of a simple model. We use the term 'metastability' instead 
of `stability' because {\it finite size} fluctuations can lead to 
violation of cylindrical symmetry and decay of a Q ring to a Q 
ball as demonstrated by our numerical simulations. 

Consider a complex scalar field whose dynamics is determined by 
the Lagrangian \be \label{model} {\cal L}={1\over 2} 
\partial_\mu \Phi^* 
\partial^\mu \Phi - U(|\Phi |) \ee
The model has a global $U(1)$ symmetry and the associated 
conserved Noether current is \be \label{current} J_\mu = Im(\Phi^* 
\partial_\mu \Phi) \ee with conserved Noether charge $ Q=\int 
d^3 x \; J_0 $.  Provided that the potential of (\ref{model}) 
satisfies certain conditions \cite{c85,lp92} the model accepts 
stable Q ball solutions which are described by the ansatz $ \Phi = 
f(r) e^{i \omega t}$.  The energy density of this Q ball 
configuration is localized and spherically symmetric. The 
stability is due to the conserved charge $Q$. 

In addition to the $Q$ ball there are other similar stable 
configurations with cylindrical or planar symmetry but infinite, 
not localized energy in three dimensions. For example an infinite 
stable Q string that extends along the z axis is described by the 
ansatz \be \label{stranz} \Phi = f(\rho) e^{i \omega t}\ee where 
$\rho$ is the azimouthal radius.  This configuration has also been 
called `planar' or 'two dimensional' Q ball\cite{lp92}. 

The energy of this configuration can be made finite and localized 
in three dimensions by considering closed Q strings. These 
configurations which have been shown to be produced during 
spherical Q ball collisions\cite{bs00,lpwww} are unstable towards 
collapse due to their tension. In order to stabilize them we need 
a pressure term that will balance the effects of tension. This 
term appears if we substitute the string ansatz (\ref{stranz}) by 
the ansatz of the form \be \label{qringanz}\Phi = f(\rho) e^{i 
\omega t} e^{i \alpha(z)} \ee where $\alpha(z)$ is a phase that 
varies uniformly along the z axis. This phase introduces a new 
non-zero $J_z$ component to the conserved current density 
(\ref{current}). The corresponding current is of the form \be 
\label{jzcons} I_z=  \int d z \;{{d \alpha}\over {d z}} \; 
2\pi\int d\rho \; \rho \; f^2 \ee  Consider now closing the 
infinite Q string ansatz (\ref{qringanz}) to a finite (but large) 
loop of size $L$. The energy of this configuration may be 
approximated by \bea E &=& {{Q^2}\over {4 \pi L \int d \rho \; 
\rho \; f^2}} + \pi \; L \; \int d \rho \; \rho \; f'^2 \nn 
\\ &+& {{(2\pi N)^2 \pi}\over L}\int d \rho \; \rho \; f^2 + 2 \pi  L \int d 
\rho \; \rho U(f)\nn \\ &\equiv & I_1 + I_2 + I_3 + I_4 \eea where 
we have assumed $\alpha (z) = {{2 \pi N}\over L} z $ and the terms 
$I_i$ are all positive. Also $Q$ is the charge conserved in $3D$ 
defined as 
\be
Q=\omega 2\pi L\int d\rho \; \rho \;f^2 \ee The winding $2 \pi 
N=\int d z \;{{d \alpha}\over {d z}}$ is topologically conserved 
and therefore the current (\ref{jzcons}) is very similar to the 
current of superconducting strings.   

After a rescaling $ \rho \longrightarrow \sqrt{\lambda_1} \rho$, $ 
z\longrightarrow \lambda_2 z $ the rescaled energy may be written 
as \be E={1 \over {\lambda_1 \lambda_2}} I_1 + \lambda_2 I_2 
+{\lambda_1 \over \lambda_2} I_3 + \lambda_1 \lambda_2 I_4 \ee 
This configuration can be metastable towards collapse since 
Derrick's theorem\cite{d64} is evaded due to the time 
dependence\cite{k97,akpf00} of the configuration (\ref{qringanz}). 
Demanding metastability towards collapse in any direction we 
obtain the virial conditions  \ba I_3 + I_4 &=& I_1 
\label{virial1} 
\\ I_2 + I_4 &=& I_1 + I_3 \label{virial2} \ea 

In order to check the validity of these conditions numerically we 
must first solve the ode which $f$ obeys. This is of the form 
\be
f'' +{1\over \rho} f' + (\omega^2 -(2\pi N)^2/L^2) f -U'(f) = 0 
\label{fode} \ee 
with boundary conditions $ f(\infty)=0$ and  ${df\over d\rho}(0)= 
0$. Equation (\ref{fode}) is identical with the corresponding 
equation for 2D Qballs\cite{akpf00} (see ansatz (\ref{stranz})) 
with the replacement of $\omega^2$ by \be \omega^2 -{{(2\pi 
N)^2}\over {L^2}} \equiv \omega'^2\ee Solutions of (\ref{fode}) 
for various $\omega'$ and $U(f) = {1 \over 2}f^2 - {1 \over 3} f^3 
+ {B \over 4}f^4$ with $B=4/9$ were obtained in Ref. 
\cite{akpf00}. Now it is easy to see that the first virial 
condition (\ref{virial1}) may be written as 
\be
\omega'^2 \int d\rho \; \rho \; f^2 = 2 \int d\rho \; \rho U(f) 
\label{virnew} \ee This is exactly the virial theorem for 2D 
Qballs (infinite Q strings) with $N=0$ and field ansatz given by 
(\ref{stranz}) with $\omega$ replaced by $\omega'$. The validity 
of this virial condition has been verified in Ref. \cite{akpf00}. 
This therefore is an effective verification of our first virial 
condition (\ref{virial1}). 

The second virial condition (\ref{virial2}) can be written (using the first virial (\ref{virial1})) as
\be
2 I_3 = I_2
\ee 
which implies that
\be
{{2\pi N^2} \over L^2} = {{\int d\rho \; \rho f'^2}\over {\int 
d\rho \; \rho f^2}} \label{vir2} \ee This can be viewed as a 
relation determining the value of $L$ required for balancing the 
tension ie for metastability.   

These virial conditions can be used to lead to a determination of 
the energy  as \be E=2(I_1+I_3)  \ee In the thin wall limit where 
$2 \pi \int d \rho \rho f^2 = A f_0^2$ ($A$ is the surface of a 
cross section of the Q ring) this may be written as \be E \simeq 
{Q^2 \over {2 L A f_0^2}} +{{(2\pi N)^2 A f_0^2}\over {2 L}} 
\label{etwa} \ee  and can be minimized with respect to $f_0^2$. 
The value of $f_0$ that minimizes the energy in the thin wall 
approximation is 
\be
f_0=\sqrt{Q\over {2\pi N A}} \ee Substituting this value back on 
the expression (\ref{etwa}) for the energy we obtain \be E={{2 \pi 
N Q}\over L} \ee This is consistent with the corresponding 
relation for spherical Q balls which in the thin wall 
approximation lead to a linear increase of the energy with $Q$. 

The above virial conditions demonstrate the persistance of the Q 
ring configuration towards shrinking or expansion in the two 
periodic directions of the Q ring torus for large radius. In order 
to study the Q rings of any size and its stability properties 
towards any type of fluctuation we must  study the full evolution 
of a Q ring in 3D by performing energy minimization and  numerical 
simulation of  evolution. This is precisely what we did for a 
potential energy given by : 
 \be 
\label{potential} U(\phi) = {1 \over 2}|\Phi|^2 - {1 \over 3} 
|\Phi|^3 + {B \over 4}|\Phi|^4 \ee The ansatz we used that 
captures the above mentioned properties of the Q ring is \be 
\label{eq:ansatz} \Phi = f(\rho,z)\; e^{i [\omega t + n\phi]} \ee 
where the center of the coordinate system now is in the center of 
the torus that describes the Q ring and the ansatz is valid for 
{\it any} radius of the Q ring. We have also replaced $N$ by $n$. 

The energy of this configuration is \ba  E  &=& {1 \over 2} {Q^2 
\over \int f^2 dV} 
  + {1 \over 2} \int \left[ \left({\partial f \over \partial \rho} \right)^2
       + {n^2 f^2 \over \rho^2} \right]\;dV \nn \\
  &+& {1 \over 2} \int 
       \left[ \left({\partial f \over \partial z} \right)^2 \right]\;dV 
       + \int U(f)\,dV  \label{eq:energy2}
  \ea The field equation for $\Phi$ is 
\begin{equation}
\label{eq:equation} \ddot{\Phi} - \Delta \Phi + \Phi - |\Phi| \Phi 
+ B |\Phi|^2 \Phi = 0 
\end{equation} Substituting the ansatz (\ref{eq:ansatz}) we find that $f(\rho,z)$ 
should satisfy 
\begin{equation}
\label{eq:ansatzequation} 
 {\partial^2 f \over \partial \rho^2} 
  + {1 \over \rho}\,{\partial f \over \partial \rho}
  - {n^2 f \over \rho^2} + {\partial^2 f \over \partial z^2}
  + (\omega^2-1) f + f^2 - B f^3 = 0
\end{equation} 
In order to solve this equation we minimized the energy 
(\ref{eq:energy2}) at fixed $Q$ using the algorithm \ba {\partial 
f \over \partial \tau} &=& - {\delta E \over \delta f} \Rightarrow  
\label{eq:dissipative}\\ {\partial f \over 
\partial \tau} &=& {\partial^2 f \over \partial \rho^2} 
  + {1 \over \rho}\,{\partial f \over \partial \rho} 
  - {n^2 f \over \rho^2} + {\partial^2 f \over \partial z^2}
  \nn \\ &+& (\omega^2-1) f  + f^2 - B f^3 \label{fevol}
\ea with boundary conditions $f(0,z)=0$, ${{\partial 
f(\rho,z)}\over  {\partial z}}|_{z=0}=0$. The validity of the 
algorithm is checked by  \begin{equation} dG/d\tau= \delta 
E/\delta f\; df/d\tau = - (dE/d\tau)^2<0 
\end{equation} In (\ref{eq:dissipative}) $\omega$ is defined as 
\be \label{omegdef}  \omega={Q \over {\int{f^2\,dV}}} \ee In the 
algorithm, we have used the initial ansatz: \begin{equation} 
f(\rho,z) = \hbox{const} \; \exp^{-{(\rho-\rho_0)^2+z^2 \over 
\hbox{const}}} 
\end{equation} where $\rho_0$ is a fixed initial radius. 
The energy minimization resulted to a non-trivial configuration 
$f(\rho,z)$ for a given set of parameters $B, n, Q$ in the 
expression for the energy. We then used (\ref{omegdef}) to 
calculate $\omega$ and constructed the full Q ring configuration 
using (\ref{eq:ansatz}).While the details of our numerical 
analysis can be found elsewhere we just report the main results.

It was verified that the Q ring configurations evolve with 
practically no distortion and are metastable despite their long 
evolution. Finite size nonsymmetric fluctuations were found to 
lead to a break up and eventual decay of the Q ring to one or more 
Q balls. Thus a Q ring is a metastable as opposed to a stable 
configuration. 

The Q ring configuration we have discovered is the simplest 
metastable ring-like defect known so far. Previous attempts to 
construct metastable ring-like configurations were based on pure 
topological arguments (Hopf maps) and required gauge fields to 
evade Derrick's theorem due to their static 
nature\cite{Faddeev:1997zj,Perivolaropoulos:2000gn}. This resulted 
in complicated models that were difficult to study analytically or 
even numerically. Q rings require only a single complex scalar 
field and they appear in all theories that admit stable Q balls 
including the minimal supersymmetric standard model (MSSM). The 
simplicity of the theory despite the non-trivial geometry of the 
field configuration is due to the combination of topological with 
non-topological charges that combine to secure metastability 
without added field complications. 

The derivation of metastability of this configuration opens up 
several interesting issues that deserve detailed investigation. 
They pertain to the various mechanisms of formation of Q Rings 
(Kibble and Affleck-Dine mechanisms, Q ball collisions etc.) as 
well as on the dependence of the winding N on Q. We hope to have 
something interesting to report in the forthcoming 70th birthday 
celebration of Holger.

\widetext 

\end{document}